\documentclass[preprint, 3p]{elsarticle}
\journal{Physics Letters B}

\usepackage{hyperref}
\usepackage{amsmath}
\usepackage{graphicx}
\usepackage{xcolor}
\usepackage{multirow}
\usepackage{bm}

\allowdisplaybreaks[4]
\newcommand{\nn}{\nonumber}
\newcommand{\beq}{\begin{equation}}
\newcommand{\eeq}{\end{equation}}
\newcommand{\bqa}{\begin{eqnarray}}
\newcommand{\eqa}{\end{eqnarray}}

\begin{document}

\title{${\mathcal O}(\alpha_s^2)$ corrections to $J/\psi+\chi_{c0,1,2}$ production at $B$ factories}


\author[a,e]{Wen-Long Sang}       \ead{wlsang@swu.edu.cn}
\author[b,c]{Feng Feng}         \ead{f.feng@outlook.com}
\author[c,d]{Yu Jia}            \ead{jiay@ihep.ac.cn}
\author[c,d]{Zhewen Mo}         \ead{mozw@ihep.ac.cn}
\author[c,d]{Jia-Yue Zhang} \ead{zhangjiayue@ihep.ac.cn}


\address[a]{School of Physical Science and Technology, Southwest University, Chongqing 400700, China}
\address[e]{College of Physics, Chongqing University, Chongqing 401331, China}
\address[b]{China University of Mining and Technology, Beijing 100083, China}
\address[c]{Institute of High Energy Physics and Theoretical Physics Center for Science Facilities, Chinese Academy of Sciences, Beijing 100049, China}
\address[d]{School of Physics, University of Chinese Academy of Sciences, Beijing 100049, China}

\begin{abstract}
We compute the ${\mathcal O}(\alpha_s^2)$ corrections 
to the exclusive channels $e^+e^-\to J/\psi+\chi_{cJ}$ ($J=0,1,2$) at $\sqrt{s}=10.58$ GeV
within the nonrelativistic QCD (NRQCD) factorization framework.
The validity of NRQCD factorization at ${\cal O}(\alpha_s^2)$ has been confirmed
for these double-charmonium exclusive production
processes.  We analyze the impact of the $\mathcal{O}(\alpha_s^2)$ corrections on the polarized and
unpolarized cross sections, as well as the $J/\psi$ angular distributions, which largely reduce the renormalization scale dependence
but increase the ${\mathcal O}(\alpha_s)$ NRQCD predictions to some extent for $\chi_{c0,1}$.
With high numerical accuracy, our predictions for $\sigma(J/\psi+\chi_{c1,2})$ through ${\mathcal O}(\alpha_s^2)$ are compatible with the upper limit of
the \texttt{Belle} measurement.
Although the theoretical prediction for $\sigma(J/\psi+\chi_{c0})$ is consistent with both
\texttt{Belle} and \texttt{BaBar} measurements within uncertainties, there still exists serious
tension between the predicted and the measured profiles for the $J/\psi$ angular distribution.
Regarding the bright observation prospect of the $e^+e^-\to J/\psi+\chi_{c1,2}$ channels in \texttt{Belle 2} experiment,
it is interesting to compare the future measurements with our NRQCD predictions.
The more accurate measurement of $e^+e^-\to J/\psi+\chi_{0}$ at \texttt{Belle 2} will also help to clarify
the long-standing puzzle of $J/\psi$ angular distribution.
\end{abstract}

\maketitle
\flushbottom

\section{Introduction}

Exclusive double charmonium production at $e^+e^-$ collider is among the simplest
hard exclusive reactions in perturbative QCD. The benchmark processes, exemplified by
$e^+e^-\to J/\psi+H$, with $H=\eta_c, \chi_{c0,1,2},\ldots$ representing a
$C$-even charmonium that recoils against $J/\psi$,
are first observed at two $B$ factories in the beginning of this century~\cite{Belle:2002tfa,BaBar:2005nic}.
This somewhat unexpected discovery has stimulated long-lasting theoretical interests,
since these types of hard exclusive reactions serve a novel and fertile playground for
unraveling the interplay between perturbative and non-perturbative aspects
in the heavy flavor sector of QCD.
In the following years, it became gradually clear that the conventional collinear factorization
approach~\cite{Lepage:1980fj,Chernyak:1983ej} is of rather limited use for these types of  processes,
especially problematic when handling the helicity-suppressed one such as $e^+e^-\to J/\psi+\eta_c$.
On the contrary, the nonrelativistic QCD (NRQCD) factorization approach~\cite{Bodwin:1994jh},
a modern effective-field-theory-based approach that explicitly exploits
the nonrelativistic nature of heavy quarkonium,  provides a much more powerful and
systematic tool kit to tackle double charmonium production processes.
NRQCD factorization approach organizes the predictions in a double expansion form, where the expansion parameters
are $v$, the typical velocity of heavy quark inside
a heavy quarkonium, and $\alpha_s$, the strong coupling constant.

To date the best studied exclusive charmonium production process is the one that
involves two $S$-wave charmonia in the final state, {\it i.e.}, $e^+e^-\to J/\psi+\eta_c$.
The severe discrepancy between initial \texttt{Belle}
measurement~\cite{Belle:2002tfa} and the lowest order NRQCD
predictions~\cite{Braaten:2002fi,Liu:2002wq,Hagiwara:2003cw} has spurred a flurry of theoretical investigations over years
(for a partial list, see Refs.~\cite{Braaten:2002fi,Liu:2002wq,Hagiwara:2003cw,Zhang:2005cha,Gong:2007db,Ma:2004qf,Bondar:2004sv,Braguta:2008tg}).
Among a variety of theoretical efforts, one key step in reconciling the discrepancy
is the discovery of a significant positive $\mathcal{O}(\alpha_s)$ correction in NRQCD factorization approach~\cite{Zhang:2005cha,Gong:2007db}.
The effect of the relativistic corrections was also carefully addressed in \cite{Bodwin:2007ga}.
The joint next-to-leading-order perturbative and relativistic correction, that is, the ${\cal O}(\alpha_s v^2)$ correction,
was also investigated in \cite{Dong:2012xx,Li:2013otv}. Recently, the very challenging $\mathcal{O}(\alpha_s^2)$ corrections to this process have also been considered in \cite{Feng:2019zmt,Huang:2022dfw}~\footnote{To appreciate the challenge
of $\mathcal{O}(\alpha_s^2)$ perturbative correction for double charmonium production processes, we quote one remark
from the 2010 review article on quarkonium physics~\cite{Brambilla:2010cs}:
"The uncalculated correction to $\sigma(e^+e^-\to J/\psi+\eta_c)$ of relative order $\alpha_s v^2$ is potentially large, as is the uncalculated correction of relative order $\alpha_s^2$. While the calculation of the former correction may be feasible, the calculation of the latter correction is probably beyond the current state of the art".}.
After piecing all kinds of available higher-order corrections together,
it appears promising that the NRQCD approach can yield a largely satisfactory description of the $B$ factory measurements for
this process, notwithstanding large uncertainties in both theory and experiment.

It is also worth noting that, the double charmonium production processes where a $P$-wave charmonium recoils
against the $J/\psi$, {\it e.g.}, $e^+e^-\to J/\psi+\chi_{cJ}$ ($J=0,1,2$),
was also reported by two $B$ factories in the early days~\cite{Belle:2002tfa,BaBar:2005nic}.
The $J/\psi+\chi_{c0}$ signals have been clearly observed,
however the $J/\psi+\chi_{c1,2}$ events, even with the more recent data set~\cite{Belle:2009bxr},
have never been established. Consequently, an upper bound has been placed on the
joint production rate $\sigma(J/\psi+\chi_{c1})+\sigma(J/\psi+\chi_{c2})$.
Fortunately, the ongoing \texttt{Belle 2} experiment, also dubbed the Super $B$ factory, 
has a designed integrated luminosity about
$50\;{\rm ab}^{-1}$, about 40 times greater than that of the terminated \texttt{Belle} experiment.
It is thus very likely that \texttt{Belle 2} experiment will be able to observe all the
$e^+e^-\to J/\psi+\chi_{cJ}$ channels with decent accuracy,
and also measure the angular distribution of the $J/\psi$.
Therefore, explaining these $S$+$P$-wave double charmonium production is interesting on its own right,
which, supplementary to the $e^+e^-\to J/\psi+\eta_{c}$ channel, constitutes a rich arena
for critically testing NRQCD factorization approach.

More than a decade ago, the $\mathcal{O}(\alpha_s)$ perturbative corrections to $e^+e^-\to
J/\psi+\chi_{cJ}$ have been computed by several groups~\cite{Zhang:2008gp, Wang:2011qg, Dong:2011fb}.
The impact of the ${\cal O}(\alpha_s)$ corrections is found to be substantial for $J/\psi+\chi_{c0}$,
albeit modest for $J/\psi+\chi_{c1,2}$. Incorporating the $\mathcal{O}(\alpha_s)$ perturbative corrections appears to be
helpful to bring the NRQCD prediction in better agreement with the measurement. Subsequently, some attempt to reduce the renormalization scale uncertainty in
these $\mathcal{O}(\alpha_s)$ corrections has also been conducted~\cite{Wang:2013vn}.
Recently, the contribution due to QED interference has also been investigated for these processes,
which has nonnegligible effect~\cite{Jiang:2018wmv}.
Furthermore, a comparative analysis of the $J/\psi$ angular distributions in a number of
double charmonium production processes has been conducted between the ${\cal O}(\alpha_s)$
NRQCD prediction and $B$ factory data~\cite{Sun:2021tma}.
It turns out that the NRQCD prediction of the $J/\psi$ angular distribution
in $e^+e^-\rightarrow J/\psi+\chi_{0}$ is in sheer contradiction to the \texttt{Belle} measurement.

Curiously, many quarkonium production and decay processes appear to suffer from substantial
higher-order radiative corrections, in particular are plagued with
large $\mathcal{O}(\alpha_s^2)$ perturbative corrections~\cite{Czarnecki:1997vz,Beneke:1997jm,Marquard:2014pea,Beneke:2014qea,Czarnecki:2001zc,Feng:2015uha,Sang:2015uxg,Feng:2017hlu}.
Stimulated by the significant one-loop radiative corrections to $e^+e^-\to J/\psi+\eta_c(\chi_{c0})$,
one naturally wonders how important the two-loop QCD corrections would be for various
exclusive double charmonium production processes.
Very recently, the two-loop QCD correction to $e^+e^-\to J/\psi+\eta_c$ has been explored for the first time~\cite{Feng:2019zmt}.
The aim of this work is to continue to calculate the $\mathcal{O}(\alpha_s^2)$ corrections to $e^+e^-\to J/\psi+\chi_{cJ}$ ($J=0,1,2$),
at lowest order in $v$. With the aid of very recent technical advancement in computing multi-loop integrals, we are able to present
the $\mathcal{O}(\alpha_s^2)$ corrections with very high numerical accuracy. In addition to assessing the impact of the
$\mathcal{O}(\alpha_s^2)$ corrections on the unpolarized cross sections,
we also consider the impact of the $\mathcal{O}(\alpha_s^2)$ corrections on the polarized cross sections as well as
the angular distributions, and confront the available \texttt{Belle} measurements.
We hope that these studies will provide important theoretical
guidance for future \texttt{Belle 2} measurements.

The rest of the paper is organized as follows.
In Sec.~\ref{sec-general-formula}, we employ the helicity amplitude formalism to analyze
the $e^- e^+ \to J/\psi + \chi_{cJ}$ ($J=0,1,2$) processes, and build the polarized and unpolarized cross sections
out of various helicity amplitudes. The angular distribution parameters
are also introduced in terms of the helicity amplitudes.
In Sec.~\ref{sec-NRQCD:fac}, we succinctly review the NRQCD factorization formula at the
helicity amplitude level, and parameterize the corresponding short-distance coefficients through $\mathcal{O}(\alpha_s^2)$.
In Sec.~\ref{Technical:strategy:calc:SDC}, we briefly describe the key technical ingredients of
extracting the short-distance coefficients affiliated with each helicity amplitude through ${\cal O}(\alpha_s^2)$.
We devote Sec.~\ref{Phenomenology} to a detailed numerical analysis of the size of the $\mathcal{O}(\alpha_s^2)$
corrections to the $e^- e^+ \to J/\psi + \chi_{cJ}$ processes, including the (un)polarized cross sections and
$J/\psi$ angular distributions. Comparison between the most refined NRQCD predictions and the existing
$B$ factory measurements is also made.
Finally, we summarize in Sec.~\ref{Summary}.

\section{(Un)polarized cross sections and angular distributions\label{sec-general-formula}}

Exclusive production of $J/\psi+\chi_{cJ}$ at $B$ factories proceeds with a simple $s$-channel spacetime route,
{\it i.e.}, that the $e^+ e^-$ first annihilate into a virtual photon, subsequently the timelike photon decays
into two charmonium final states (the latter is also referred to as the time-like electromagnetic form factor of charmonium).
Assume the $e^-$ and $e^+$ move along the $\hat{z}$ direction. We work in the center-of-mass (CM) frame with the CM energy $\sqrt{s}$.
Let $\theta$ denote the polar angle between the direction of the outgoing $J/\psi$ and the $e^-$ beam.
It is convenient to express the differential $J/\psi+\chi_{cJ}$ production rates in terms
of the differential decay rates of a time-like photon.
Moreover, to retain as much polarization information of the outgoing charmonia as possible,
it is advantageous to employ the {\it helicity amplitude formalism} following \cite{Dong:2011fb}.
Let $\lambda_1$ and $\lambda_2$ represent the helicities of the $J/\psi$ and $\chi_{cJ}$, respectively.
The differential rate of a timelike photon decay into $J/\psi(\lambda_1)+\chi_{cJ}(\lambda_2)$ becomes
\beq
\label{Polarized:Diff:Decay:Rate}
\frac{d\Gamma\left[\gamma^*(S_z)\to J/\psi(\lambda_1)+\chi_{cJ}(\lambda_2)\right]}{d\cos\theta}
  =\dfrac{|{\bf P}|}{16\pi s} \left|d_{S_z,\lambda}^1(\theta)\right|^2
  |{\mathcal A}_{\lambda_1,\lambda_2}^J|^2,
\eeq
with $\lambda\equiv\lambda_1-\lambda_2$. $S_z$ signifies the magnetic number of the timelike photon
with the $+\hat{z}$ direction as the spin quantization axis.
${\mathcal A}_{\lambda_1,\lambda_2}^J$ is the intended helicity amplitude that encapsulates
all nontrivial strong interaction dynamics,
which depends upon $\lambda_1$ and $\lambda_2$ yet not upon $S_z$.
The angular distribution is solely governed by the Wigner function $d^j_{m_1,m_2}(\theta)$ which depends upon the
quantum number $S_z$ and $\lambda$.
Note that angular momentum conservation enforces $|\lambda|\leq 1$.
$|{\bf P}|$ in \eqref{Polarized:Diff:Decay:Rate} signifies the magnitude of the 3-momentum of the
$J/\psi$ ($\chi_{cJ}$) in the CM frame:
\beq
|{\bf P}|= \sqrt{\dfrac{\lambda(s,M_{J/\psi}^2,M_{\chi_{cJ}}^2)}{4 s}},
\eeq
where $\lambda(x,y,z)=x^2+y^2+z^2-2xy-2xz-2yz$ is the K\"allen function.

Parity invariance relates various helicity amplitudes with opposite helicities of $J/\psi$ and $\chi_{cJ}$:
\beq
{\cal A}^J_{\lambda_1,\lambda_2} = (-)^J {\cal A}^J_{-\lambda_1,-\lambda_2}.
\eeq
Consequently, as constrained by angular momentum conservation,
there are left with only 2, 3 and 5 {\it independent} helicity amplitudes for a timelike photon decay
into $J/\psi+\chi_{c0}$, $J/\psi+\chi_{c1}$ and $J/\psi+\chi_{c2}$, respectively. Note that parity
invariance enforces $A^1_{0,0}=0$.

For a definite helicity configuration of $J/\psi$ and $\chi_{cJ}$,
it is straightforward to covert
the differential decay rate of a timelike photon into the differential polarized cross section in $e^+e^-$
annihilation~\cite{Dong:2011fb}:
\begin{eqnarray}
\label{Diff:Pol:X:Section:Angular:Distribution}
&&\frac{d\sigma\left[e^+e^-\to J/\psi(\lambda_1)+\chi_{cJ}(\lambda_2)\right]}{d\cos\theta}= \frac{2\pi\alpha}{s^{3/2}}
\sum_{S_z=\pm1} \frac{d\Gamma\left[\gamma^*(S_z)\to J/\psi(\lambda_1)+\chi_{cJ}(\lambda_2)\right]}{d\cos\theta}
\\
  && \qquad =\dfrac{\alpha}{8 s^2} \left(\dfrac{|{\bf P}|}{\sqrt{s}}\right)
|{\mathcal A}_{\lambda_1,\lambda_2}^J|^2\times \Bigg\{
\begin{array}{c}
\dfrac{1+\cos^2\theta}{2},\qquad\lambda={\pm 1}
\nn\\
\nn\\
1-\cos^2\theta,\qquad \lambda=0,
\end{array}
\end{eqnarray}
where $\alpha$ signifies the QED fine structure constant.
We have averaged upon the polarizations of the $e^-$ and $e^+$. Since helicity conservation in QED warrants the virtual photon
must be transversely polarized, we only need sum over two transverse polarization of the timelike photon.
It is this partial sum that is responsible for the anisotropic angular distribution patterns in \eqref{Diff:Pol:X:Section:Angular:Distribution}.
Note that regardless of the helicity configurations of two outgoing charmonia,
the angular distributions of $J/\psi$ always fit in the pattern $\propto 1\pm \cos^2\theta$.

So far $B$ factories only measured the unpolarized double charmonium production cross sections.
Summing over all possible helicities of $J/\psi$ and $\chi_{cJ}$ in \eqref{Diff:Pol:X:Section:Angular:Distribution},
we are ready to obtain the differential unpolarized  $J/\psi+\chi_{cJ}$ production rates,
which can be generically parameterized as
\beq
\label{Unpolar:Diff:X:Section:Parametrization}
\dfrac{d\sigma(e^+e^-\to J/\psi+\chi_{cJ})}{d\cos\theta} = A_J \left(1+\alpha_J\cos^2\theta\right),
\qquad J=0,1,2
\eeq
where $\alpha_J$ is a dimensionless parameter that governs the profile of the
angular distribution, subject to the constraint $|\alpha_J|\leq 1$.

After some simple massage from \eqref{Diff:Pol:X:Section:Angular:Distribution},
we obtain
\beq
\label{Unpolar:Diff:X:Section:Parametrization:chic0}
A_0=\dfrac{\alpha}{8s^2} \left(\dfrac{|{\bf P}|}{\sqrt{s}}\right)
\left\{ |{\mathcal A}_{0,0}^0|^2+ |{\mathcal A}_{1,0}^0|^2\right\},\qquad\:
\alpha_0= -\dfrac{|{\mathcal A}_{0,0}^0|^2 -|{\mathcal A}_{1,0}^0|^2}{|{\mathcal A}_{0,0}^0|^2 +|{\mathcal A}_{1,0}^0|^2}
\eeq
for $J/\psi+\chi_{c0}$,
\beq
\label{Unpolar:Diff:X:Section:Parametrization:chic1}
A_1 =\dfrac{\alpha}{8s^2} \left(\dfrac{|{\bf P}|}{\sqrt{s}}\right)
\left\{ |{\mathcal A}_{1,0}^1|^2+|{\mathcal A}_{0,1}^1|^2 + 2|{\mathcal A}_{1,1}^1|^2  \right\},
\qquad\:
\alpha_1= { |{\mathcal A}_{1,0}^1|^2+|{\mathcal A}_{0,1}^1|^2 -2|{\mathcal A}_{1,1}^1|^2
\over  |{\mathcal A}_{1,0}^1|^2+|{\mathcal A}_{0,1}^1|^2 + 2|{\mathcal A}_{1,1}^1|^2 },
\eeq
for $J/\psi+\chi_{c1}$,
\begin{subequations}
\label{Unpolar:Diff:X:Section:Parametrization:chic2}
\begin{eqnarray}
& & A_2 = {\alpha\over 8s^2} \left({|{\bf P}|\over \sqrt{s}}\right)
\left\{ |{\mathcal A}_{0,0}^2|^2 +|{\mathcal A}_{1,0}^2|^2+ |{\mathcal A}_{0,1}^2|^2 +2|{\mathcal A}_{1,1}^2|^2 + |{\mathcal A}_{1,2}^2|^2    \right\},
\\
&& \alpha_2= - { |{\mathcal A}_{0,0}^2|^2 -|{\mathcal A}_{1,0}^2|^2 - |{\mathcal A}_{0,1}^2|^2  + 2|{\mathcal A}_{1,1}^2|^2 - |{\mathcal A}_{1,2}^2|^2
\over  |{\mathcal A}_{0,0}^2|^2 +|{\mathcal A}_{1,0}^2|^2+ |{\mathcal A}_{0,1}^2|^2 +2|{\mathcal A}_{1,1}^2|^2 + |{\mathcal A}_{1,2}^2|^2   }
\end{eqnarray}
\end{subequations}
for $J/\psi+\chi_{c2}$.

Integrating \eqref{Unpolar:Diff:X:Section:Parametrization} over the polar angle,
one finds that the total unpolarized cross sections to be $\sigma(J/\psi+\chi_{cJ})=2 A_J (1+\alpha_J/3)$.
Substituting the expressions of $A_J$ and $\alpha_J$ in
\eqref{Unpolar:Diff:X:Section:Parametrization:chic0},
\eqref{Unpolar:Diff:X:Section:Parametrization:chic1} and \eqref{Unpolar:Diff:X:Section:Parametrization:chic2},
we finally arrive at
\begin{subequations}
\label{Total:Unpol:X:section:ChicJ}
\begin{eqnarray}
\sigma(J/\psi+\chi_{c0})&=&\frac{\alpha}{6s^2}\frac{|{\bf P}|}{\sqrt{s}}\bigg(|{\mathcal A}_{0,0}^0|^2 + 2|{\mathcal A}_{1,0}^0|^2\bigg),
\\
\sigma(J/\psi+\chi_{c1})&=&
\frac{\alpha}{6s^2}\frac{|{\bf P}|}{\sqrt{s}}\bigg( 2|{\mathcal A}_{1,0}^1|^2+2|{\mathcal A}_{0,1}^1|^2 +2|{\mathcal A}_{1,1}^1|^2\bigg),
\\
\sigma(J/\psi+\chi_{c2})&=&
\frac{\alpha}{6s^2}\frac{|{\bf P}|}{\sqrt{s}}\bigg(  |{\mathcal A}_{0,0}^2|^2 +2|{\mathcal A}_{1,0}^2|^2+
2|{\mathcal A}_{0,1}^2|^2 +2|{\mathcal A}_{1,1}^2|^2 + 2|{\mathcal A}_{1,2}^2|^2
\bigg).
\end{eqnarray}
\end{subequations}

The central task of this work is then to compute all the 10 helicity amplitudes ${\mathcal A}^J_{\lambda_1,\lambda_2}$,
which are functions of several entangled energy scales: $\sqrt{s}$, $m_c$, and $\Lambda_{\rm QCD}$.
Nevertheless, in the high energy limit, {\it e.g.}, $\sqrt{s}\gg m_c$, each helicity amplitude
obeys definite power law scaling~\cite{Chernyak:1980dj,Brodsky:1981kj},
$A_{\lambda_1,\lambda_2}^J \propto s^{-{1\over 2}(1+|\lambda_1+\lambda_2|)}$.
Therefore, one anticipates that the double charmonium cross section for
any prescribed helicity configuration
should exhibit the asymptotical scaling behavior~\cite{Dong:2011fb}:
\beq
\label{Polarized:X:Section:Helicity:scaling}
\sigma(J/\psi(\lambda_1)+\chi_{cJ}(\lambda_2))\propto s^{-3-|\lambda_1+\lambda_2|}.
\eeq

A direct consequence of (\ref{Polarized:X:Section:Helicity:scaling}) is that, the polarized cross section that
exhibits slowest asymptotic decrease, $\sigma\propto 1/s^3$, is the one with $|\lambda_1+\lambda_2|=0$,
which corresponds to the unique helicity configuration $(\lambda_1,\lambda_2)=(0,0)$ by angular momentum conservation.
Therefore, at asymptotic high energy limit, the total cross section of double charmonium production is saturated by
the $(0,0)$ helicity configuration~\footnote{An interesting exception is for the channel $J/\psi+\chi_{c1}$.
Since the $(0,0)$ configuration is forbidden by parity, the leading contributions arise from either $(1,0)$ or $(0,1)$
channels, thus, according to \eqref{Polarized:X:Section:Helicity:scaling}, we anticipate that
the cross section for $J/\psi+\chi_{c1}$ is suppressed by an extra power of $1/s$
with respect to $J/\psi+\chi_{c0,2}$.}. Since the $B$ factory energy is not much bigger than the charmonium mass,
one should not be too surprised if the hierarchy of different polarized cross sections from our actual calculation
differs considerably from what is  anticipated in \eqref{Polarized:X:Section:Helicity:scaling}.

After incorporating higher-order perturbative corrections, the scaling behavior in
\eqref{Polarized:X:Section:Helicity:scaling} still holds,
yet up to mild logarithmic violation (modulo powers of $\ln s$).

\section{NRQCD factorization of the $\gamma^*\to J/\psi+\chi_{cJ}$ helicity amplitude\label{sec-NRQCD:fac}}

NRQCD factorization not only holds for inclusive quarkonium production processes, but also
for hard exclusive quarkonium production processes.
For $\gamma^*\to J/\psi+\chi_{cJ}$ process, NRQCD factorization is also applicable at helicity amplitude level.
Concretely speaking, at lowest order in $v$, the helicity amplitude for timelike photon decay into $J/\psi(\lambda_1)+\chi_{cJ}(\lambda_2)$
can be put in a factorized form:
\begin{eqnarray}
\label{NRQCD:factorization:helicity:ampl}
{\mathcal A}_{\lambda_1,\lambda_2}^J =  {\mathcal C}^J_{\lambda_1,\lambda_2}(s, m_c^2, \mu^2_\Lambda)\,
{ \langle \mathcal{O}_{{}^3S_1}(\mu_\Lambda)\rangle \langle{\cal O}_{^3P_J}(\mu_\Lambda)\rangle \over m_c^3}.
\end{eqnarray}
$\mathcal{C}^J_{\lambda_1,\lambda_2}$ in \eqref{NRQCD:factorization:helicity:ampl} signifies the {\it dimensionless}
short-distance coefficient (SDC) for each corresponding helicity amplitude. $\mu_\Lambda$ denotes NRQCD factorization scale, which
enters both the SDC and the nonperturbative vacuum-to-charmonium NRQCD matrix element, and
\begin{subequations}
\label{NRQCD:matrix:elements}
\begin{eqnarray}
\langle \mathcal{O}_{{}^3S_1}(\mu_\Lambda)\rangle &=&\langle J/\psi|
\psi^{\dagger} {\bm\sigma}\cdot {\bm\varepsilon}_{J/\psi} \chi(\mu_\Lambda) |0\rangle\\
\langle{\cal O}_{^3P_0}(\mu_\Lambda)\rangle&=&\left\langle \chi_{cJ}\left|
\psi^{\dagger}\frac{1}{\sqrt{3}}\left(-\frac{i}{2}\overleftrightarrow{{\bf
D}}\cdot {\bm\sigma}\right) \chi(\mu_\Lambda)\right|0\right\rangle,
\\
\langle{\cal O}_{^3P_1}(\mu_\Lambda)\rangle&=&\left\langle \chi_{cJ}\left|
\psi^{\dagger}\frac{1}{\sqrt{2}}\left(-\frac{i}{2}\overleftrightarrow{{\bf
D}}\times\bm{\sigma}\right)\cdot  {\bm\varepsilon}_{\chi_{c1}}\chi(\mu_\Lambda) \right|0\right\rangle,
\\
\langle{\cal O}_{^3P_2}(\mu_\Lambda)\rangle&=&\left\langle \chi_{cJ}\left|
\psi^{\dagger}\left(-{i\over 2}\overleftrightarrow{D}^{(i}\sigma^{j)} \varepsilon_{\chi_{c2}}^{ij}\right)\chi(\mu_\Lambda) 
\right|0\right\rangle,
\end{eqnarray}
\end{subequations}
where 
$\psi^\dagger$ and $\chi$ in \eqref{NRQCD:matrix:elements} denote the Pauli spinor fields creating a charm quark and anticharm quark,
$\bm{\varepsilon}_{J/\psi}$ denotes the polarization vector of $J/\psi$ in the rest frame, $\varepsilon_{\chi_{c1}}$ ($\varepsilon_{\chi_{c2}}$) represent the
polarization vector (tensor) for $\chi_{c1}$ ($\chi_{c2}$) at rest.

In phenomenological analysis, those long-distance NRQCD matrix elements occurring in \eqref{NRQCD:factorization:helicity:ampl}
are often approximated by the radial Schr\"odinger wave functions at the origin ($J/\psi$) and the first derivative of the $P$-wave radial wave functions at the origin ($\chi_{cJ}$):
 \begin{subequations}
\label{NRQCD:vac:to:onium:ME:wvfn:at:origin}
\begin{eqnarray}
\langle{\cal O}_{^3S_1}(\mu_\Lambda\approx 1\;\mathrm{GeV})\rangle &\approx& \sqrt{ N_c\over 2\pi}{R_{J/\psi}}(0),
\\
\langle{\cal O}_{^3P_J}(\mu_\Lambda\approx 1\;\mathrm{GeV})\rangle&\approx& \sqrt{3 N_c\over 2\pi} {R^\prime_{\chi_{cJ}}}(0),
\end{eqnarray}
\end{subequations}
where $N_c=3$ denotes the number of color, we have also tacitly assumed
${R^\prime_{\chi_{c0}}}(0)\approx {R^\prime_{\chi_{c1}}}(0) \approx {R^\prime_{\chi_{c2}}}(0)$
by appealing to approximate heavy quark spin symmetry. We stress that the heavy quark spin symmetry breaking effect is of relative order-$v^2$, therefore ${R^\prime_{\chi_{c0,1,2}}}(0)$ are expected to differ $30\%$ from each other. Due to renormalization effect, these phenomenological (derivative of) wave functions
at the origin in quark potential model must be promoted as scale-dependent NRQCD matrix elements.

Through ${\cal O}(\alpha_s^2)$, the SDC associated with each helicity amplitude
is expected to take the following structure:
\begin{eqnarray}
\label{SDCs:power:series:in:alphas}
&& \mathcal{C}_{\lambda_1,\lambda_2}^J \left( r, {\mu_R^2\over m_c^2}, {\mu_\Lambda^2\over m_c^2} \right)
= \dfrac{64\pi e\alpha_s}{27\sqrt{3}} \: r^{(1+|\lambda_1+\lambda_2|)/2} \: \mathcal{C}_{\lambda_1,\lambda_2}^{J(\rm tree)}\bigg\{1+\frac{\alpha_s(\mu_R)}{\pi}
\bigg(\frac{1}{4}\beta_0\ln\frac{\mu_R^2}{m_c^2}+ c_{\lambda_1,\lambda_2}^{J(1)}\bigg)
\nn\\
&&+ \frac{\alpha_s^2(\mu_R)}{\pi^2}\bigg(\frac{1}{16}\beta_0^2\ln^2\frac{\mu_R^2}{m_c^2}+\frac{1}{16}
(8 c_{\lambda_1,\lambda_2}^{J(1)}\beta_0+\beta_1)\ln\frac{\mu_R^2}{m_c^2}+(\gamma_{{}^3S_1}+\gamma_{{}^3P_J})
\ln\frac{\mu_\Lambda^2}{m_c^2}+ c_{\lambda_1,\lambda_2}^{J(2)}\bigg)\bigg\},
\nn\\
\end{eqnarray}
where $r\equiv 4m_c^2/s$ is a dimensionless ratio, $\mu_R$ and $\mu_\Lambda$ refer to the renormalization scale and NRQCD factorization scale, respectively.
$\beta_0=11C_A/3-4T_F n/3$ and $\beta_1=34C_A^2/3-20C_A T_F n/3-4C_F T_F n$ (with $T_F=1/2$, $C_F=(N_c^2-1)/2N_c$, $C_A=N_c$)
are the first two coefficients in the QCD $\beta$ function.
$n$ signifies he number of the active flavors $n=n_H+n_L$, with the number of light quarks
$n_L=3$, and the number of heavy quarks $n_H=2$~\footnote{Since the $B$ factory energy exceeds 
twice bottom quark mass, we explicitly include the $b$ quark contribution inside the loop diagrams.}.
Note that the occurrence of $\ln \mu_R$ terms guarantees the renormalization group invariance of the
$\mathcal{C}_{\lambda_1,\lambda_2}^J$ at two-loop accuracy. We have deliberately pulled out the $r$-dependent factor in front to make 
the helicity scaling rule manifest, 
so that the SDCs $\mathcal{C}_{\lambda_1,\lambda_2}^{J(\rm tree)}$ scale as $r^0$.

The tree-level SDCs $\mathcal{C}_{\lambda_1,\lambda_2}^{J(\rm tree)}$ have been known long ago~\cite{Braaten:2002fi, Dong:2011fb}:
\begin{subequations}
\label{SDC:helicity:amplitude:tree}
\begin{eqnarray}
\mathcal{C}_{0,0}^{0({\rm tree})}&=&1+10 r-12 r^2,\qquad
\mathcal{C}_{1,0}^{0({\rm tree})}={9-14 r},
\\
\mathcal{C}_{1,0}^{1({\rm tree})}&=&{-\sqrt{6} r},\qquad
\mathcal{C}_{0,1}^{1({\rm tree})}= -{\sqrt{6} (2-7r)},\qquad
\mathcal{C}_{1,1}^{1({\rm tree})}=  -{2\sqrt{6} (1-3r)},
\\
\mathcal{C}_{0,0}^{2({\rm tree})}&=&-\sqrt{2}(1-2 r-12 r^2), \qquad \mathcal{C}_{1,0}^{2({\rm tree})}= -\sqrt{2} (3-11r),
\\
\mathcal{C}_{1,1}^{2({\rm tree})}&=&-2\sqrt{6} (1-3r),\qquad
\mathcal{C}_{0,1}^{2({\rm tree})}=-\sqrt{6} (1-5r) , \qquad \mathcal{C}_{1,2}^{2({\rm tree})}= -2\sqrt{3}.
\end{eqnarray}
\end{subequations}
Accidently, the helicity amplitudes $\gamma^*\to J/\psi(\pm 1)+\chi_{c1}(0)$ are more suppressed than as suggested
from the helicity scaling rule.

The coefficients $c_{\lambda_1,\lambda_2}^{J (1)}$ in \eqref{SDCs:power:series:in:alphas} encode
the ${\cal O}(\alpha_s)$ corrections to each of 10 helicity amplitudes, which were first computed in \cite{Dong:2011fb}.
The exact expressions are somewhat cumbersome to be presented in the text, yet the asymptotic ones of  $c_{\lambda_1,\lambda_2}^{J (1)}$
become quite succinct, whose analytic form can also be found in \cite{Dong:2011fb}. A noteworthy fact is that,
the helicity-conserving $(0,0)$ channels are always accompanied by the single logarithm $\ln s/m_c^2$,
while the helicity-suppressed channels are always accompanied by the double logarithm $\ln^2 s/m_c^2$.

The central theme of this work is to compute $c_{\lambda_1,\lambda_2}^{J(2)}$,
the two-loop perturbative corrections to the SDC $\mathcal{C}_{\lambda_1,\lambda_2}^J$.
If the NRQCD factorization remains valid  for this exclusive double charmonium production,
we expect that the SDC should develop
a logarithmic dependence on NRQCD factorization scale $\mu_\Lambda$ start at ${\cal O}(\alpha_s^2)$,
with the form as exactly prescribed in \eqref{SDCs:power:series:in:alphas}.
The coefficients $\gamma_{J/\psi}$ and $\gamma_{\chi_{c0}}$ signify the
anomalous dimensions associated with the NRQCD
bilinear operators carrying the quantum numbers of $^3S_1$ and $^3P_J$, which are defined through
 \begin{align}
        \dfrac{\mathrm{d} \ln\langle\mathcal{O}_{{}^3L_J}(\mu_\Lambda)\rangle}{\mathrm{d}\ln\mu_\Lambda^2}=-\left(\dfrac{\alpha_s(\mu_\Lambda)}{\pi}\right)^2\gamma_{{}^3L_J}+\mathcal{O}\left(\alpha_s^3\right),\label{eq-RG}
\end{align}
and have already been known from various sources~\cite{Czarnecki:1997vz,Beneke:1997jm,Hoang:2006ty,Sang:2015uxg,Sang:2020fql}:
\begin{subequations}\label{eq-snomalous-dimension}
\begin{eqnarray}
\gamma_{{}^3S_1}&=&-\pi^2\bigg(\frac{C_AC_F}{4}+\frac{C_F^2}{6}\bigg),\\
\gamma_{{}^3P_0}&=&-\pi^2\bigg(\frac{C_AC_F}{12}+\frac{C_F^2}{3}\bigg),\\
\gamma_{{}^3P_1}&=&-\pi^2\bigg(\frac{C_AC_F}{12}+\frac{5C_F^2}{24}\bigg),\\
\gamma_{{}^3P_2}&=&-\pi^2\bigg(\frac{C_AC_F}{12}+\frac{13C_F^2}{120}\bigg).
\end{eqnarray}
\end{subequations}

\section{Description of the steps to deduce NRQCD SDCS\label{Technical:strategy:calc:SDC}}
\begin{figure}[tbh]
  \begin{center}
    \includegraphics[width=0.9\textwidth]{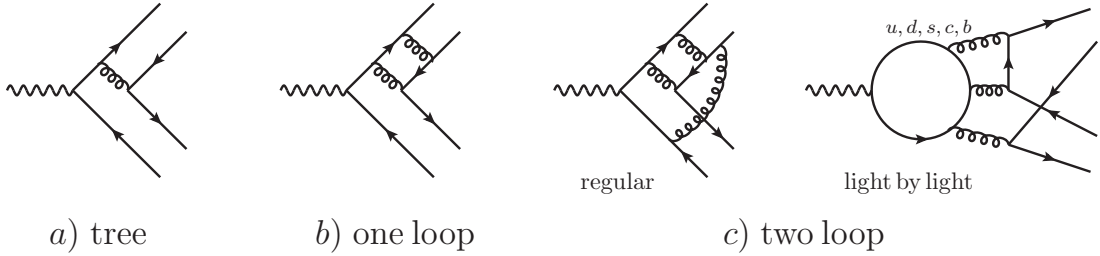}
    \caption{Representative diagrams for $\gamma^*\to c\bar{c}({}^3S_1^{(1)})+c\bar{c}({}^3P_J^{(1)})$
    through two-loop order.
    \label{feynman:diagrams}}
  \end{center}
\end{figure}

In principle the SDCs $\mathcal{C}_{\lambda_1,\lambda_2}^J$ can be inferred by the perturbative matching procedure,
{\it i.e.}, by replacing the physical $J/\psi$ and $\chi_{cJ}$ with the
fictitious quarkonia composed of the free $c\bar{c}$ pairs with quantum numbers ${}^3S_1^{(1)}$ and ${}^3P_J^{(1)}$,
and computing both sides in \eqref{NRQCD:factorization:helicity:ampl} in perturbative QCD and NRQCD,
then solving for the SDCs order by order in perturbation theory~\footnote{For a rigorous perturbative matching calculation for $P$-wave quarkonium exclusive production,
we refer the interested readers to \cite{Brambilla:2017kgw} that computes the leading relativistic corrections to $e^+e^-\to \chi_{cJ}+\gamma$.}.
Nevertheless, since here we are only concerned with the lowest order in velocity expansion, practically it is more efficient
to utilize the well-known covariant color/spin/orbital projector technique to expedite the matching calculation~\cite{Petrelli:1997ge}.
This covariant projector method is a standard tool in computing quarkonium production and decay, which has been previously employed to compute the LO
and $\mathcal{O}(\alpha_s)$ perturbative contributions to the $e^+e^-\to J/\psi+\chi_{cJ}$ processes~\cite{Braaten:2002fi,Dong:2011fb}.
In this work, we also employ this shortcut to project out the intended quark amplitude
$\gamma^*\to c\bar{c}({}^3S_1^{(1)})+c\bar{c}({}^3P_J^{(1)})$.
To further extract all ten helicity amplitudes, we find it convenient to employ various covariant helicity projectors constructed in \cite{Xu:2012uh}.

Through this work we adopt the Feynman gauge and use dimensional regularization to regularize the occurring UV and IR divergences.
We apply the packages \texttt{QGraf}~\cite{Nogueira:1991ex} and \texttt{FeynArts}~\cite{Hahn:2000kx} to generate the corresponding
Feynman diagrams and amplitudes for $\gamma^*\to c\bar{c}c\bar{c}$ through $\alpha_s^2$.  We implement the packages \texttt{FeynCalc}/\texttt{FormLink}~\cite{Mertig:1990an,Feng:2012tk} to handle
the Lorentz index contraction and Dirac/$SU(N_c)$ traces, in order to utilize
the spin/color/orbital/helicity projectors to project out the respective helicity amplitude for $\gamma^*\to c\bar{c}({}^3S_1^{(1)})+c\bar{c}({}^3P_J^{(1)})$.
We pause to emphasize that, in applying the covariant projection method, we have tacitly expanded the QCD amplitudes
in powers of quark relative momentum prior to conducting loop integrals.
This amounts to directly extracting the NRQCD SDCs in the context of method of region~\cite{Beneke:1997zp}, which is considerably simpler than
the literal matching calculation.

Nearly 2000 two-loop diagrams survive for the $\gamma^*\to c\bar{c}({}^3S_1^{(1)})+c\bar{c}({}^3P_J^{(1)})$ processes.
Some representative  Feynman diagrams through $\alpha_s^2$ are displayed  in Figure~\ref{feynman:diagrams}. The two-loop diagrams can be divided into two categories, the regular part and ``light-by-light" part. The latter admits a peculiar topology where a closed quark loop is linked with the timelike photon and three gluons. Since the sum of electric charge of the light flavors cancels, {\it e.g.}, $e_u+e_d+e_s=0$, so we can simply ignore the ``light-by-light" contributions stemming from the light quark loops.

Employing the package \texttt{Apart}~\cite{Feng:2012iq} for partial fractions and \texttt{FIRE}~\cite{Smirnov:2014hma}
for integration-by-parts (IBP) reduction, we end up with roughly 600 two-loop master integrals (MIs).
The biggest challenge of this work is to precisely compute these MIs, many of which bears rather complicated topology and are generally complex-valued.
It turns out that it becomes a formidable task for the traditional numerical recipes such as sector decomposition to
yield satisfactory results. Fortunately, a powerful new algorithm dubbed {\it Auxiliary Mass Flow} (AMF) has recently been pioneered by Liu and Ma~\cite{Liu:2017jxz,Liu:2020kpc,Liu:2021wks,Liu:2022mfb}.
The key idea is to set up differential equations with respect to an auxiliary mass variable, with the vacuum bubble diagrams as the
boundary conditions. Remarkably, the differential equations can be solved iteratively with very high numerical precision in much shorter time.
We have examined that, the AMF approach can readily tackle all the complex-valued MIs with high precision, which turns out to be much more superior to
the sector decomposition method for the MIs encountered in this work.

We thus employ the newly released package~\texttt{AMFlow}~\cite{Liu:2022chg} to compute all the 600 MIs.
After implementing the charm quark mass and field strength on-shell renormalization, and renormalizing the QCD coupling
under $\overline{\rm MS}$ scheme, we numerically verify that all the UV poles indeed cancel, yet each renormalized quark helicity amplitude is left with
a single IR pole, whose coefficients are exactly identical to $(\gamma_{{}^3S_1}+\gamma_{{}^3P_J})/2$ in \eqref{SDCs:power:series:in:alphas}.
This can be viewed as a highly nontrivial success of NRQCD factorization for exclusive double charmonium production at two loop order.
We can factor these IR divergences using $\overline{\rm MS}$ prescription, so that the two-loop SDC in each helicity channel starts to
develop an explicit logarithmic dependance on NRQCD factorization scale $\mu_\Lambda$. It is straightforward to infer the UV/IR finite
non-logarithmic part of the two-loop SDC, $c_{\lambda_1,\lambda_2}^{J(2)}$.

\section{Phenomenology\label{Phenomenology}}

\renewcommand\arraystretch{1.25}

\begin{table}[!htbp]\scriptsize
\caption{One-loop and two-loop contributions to the dimensionless SDCs, $c_{\lambda_1,\lambda_2}^{J (1,2)}$
in \eqref{SDCs:power:series:in:alphas}.
\label{Table-SDC:C:one:two:loop}
}
\setlength{\tabcolsep}{0.4mm}
\centering
  \resizebox{\textwidth}{!}{
  \begin{tabular}{|c|c|c|c|}
    \hline
      $H$ & $(\lambda_1,\lambda_2)$ &  $c_{\lambda_1,\lambda_2}^{(1)}$ & $c_{\lambda_1,\lambda_2}^{(2)}$\\

    \hline
  \multirow{4}*{$\chi_{c0}$}
    & \multirow{2}*{$(1,0)$} 
    & $0.6315+1.1076 i+(0.0784-0.5236 i) n_L$ 
    & $-34.73+9.11 i+(-0.8042+0.2182 i) n_L+(-0.2687-0.0821 i) n_L^2$    \\
    &                        &                                         & $+(-0.2046+0.1862 i) \text{lbl}_c+(-0.0249+0.1891 i) \text{lbl}_b$ \\
  \cline{2-4}
    & \multirow{2}*{$(0,0)$} & $0.296+2.347 i+(0.0412-0.5236 i) n_L$   
    & $-46.10+19.02 i+(0.3140-0.2819 i) n_L+(-0.2765-0.0432 i) n_L^2$    \\
    &                        &                                         & $+(-0.2688+0.2705 i) \text{lbl}_c+(-0.0394+0.1948 i) \text{lbl}_b$ \\
\hline
  \multirow{6}*{$\chi_{c1}$}
    & \multirow{2}*{$(1,1)$} & $-6.433+5.916 i+(0.0333-0.5236 i) n_L$  & $-81.86-46.54 i+(3.924+5.361 i) n_L+(-0.2781-0.0349 i) n_L^2$      \\
    &                        &                                         & $+(0.2566-0.0907 i) \text{lbl}_c+(0.4657+0.0763 i) \text{lbl}_b$   \\
  \cline{2-4}
    & \multirow{2}*{$(1,0)$} & $-46.03+56.94 i+(0.1046-0.5236 i) n_L$  & $-626.7-416.2 i+(56.22+25.82 i) n_L+(-0.2632-0.1095 i) n_L^2$      \\
    &                        &                                         & $+(0.2721+0.9668 i) \text{lbl}_c+(2.755+1.772 i) \text{lbl}_b$     \\
  \cline{2-4}
    & \multirow{2}*{$(0,1)$} & $-2.999+2.935 i+(0.0277-0.5236 i) n_L$  & $-48.57-18.06 i+(0.814+2.847 i) n_L+(-0.2793-0.0290 i) n_L^2$      \\
    &                        &                                         & $+(0.1354-0.0549 i) \text{lbl}_c+(0.2120-0.0063 i) \text{lbl}_b$   \\
\hline
  \multirow{10}*{$\chi_{c2}$}
    & \multirow{2}*{$(1,2)$} & $-3.881+5.707 i+(0.1046-0.5236 i) n_L$  & $-49.95-11.46 i+(4.387+2.203 i) n_L+(-0.2632-0.1095 i) n_L^2$      \\
    &                        &                                         & $+(-0.7295+0.3095 i) \text{lbl}_c+(-0.5207+0.3913 i) \text{lbl}_b$ \\
  \cline{2-4}
    & \multirow{2}*{$(1,1)$} & $-3.467+4.851 i+(0.0333-0.5236 i) n_L$  & $-72.48-1.12 i+(3.243+2.181 i) n_L+(-0.2781-0.0349 i) n_L^2$       \\
    &                        &                                         & $+(-0.2593+0.1258 i) \text{lbl}_c+(-0.1398+0.1887 i) \text{lbl}_b$ \\
  \cline{2-4}
    & \multirow{2}*{$(1,0)$} & $-3.222+3.459 i+(-0.0006-0.5236 i) n_L$ & $-59.64-9.29 i+(2.137+2.167 i) n_L+(-0.2852+0.0006 i) n_L^2$       \\
    &                        &                                         & $+(-0.2415+0.1598 i) \text{lbl}_c+(-0.1002+0.1359 i) \text{lbl}_b$ \\
  \cline{2-4}
    & \multirow{2}*{$(0,1)$} & $-2.897+2.256 i+(-0.0960-0.5236 i) n_L$ & $-19.35-22.71 i+(1.512+1.247 i) n_L+(-0.3052+0.1005 i) n_L^2$      \\
    &                        &                                         & $+(0.1925+0.0292 i) \text{lbl}_c+(0.2658-0.1131 i) \text{lbl}_b$   \\
  \cline{2-4}
    & \multirow{2}*{$(0,0)$} & $-4.411+4.620 i+(-0.0721-0.5236 i) n_L$ & $-66.01-27.27 i+(4.373+1.358 i) n_L+(-0.3002+0.0755 i) n_L^2$      \\
    &                        &                                         & $+(-0.1245+0.2091 i) \text{lbl}_c+(0.02997-0.00909 i) \text{lbl}_b$\\
\hline
  \end{tabular}
  }
  
\end{table}

We proceed to present our numerical predictions accurate to relative ${\cal O}(\alpha_s^2)$ and confront
the available $B$ factory data. We take the following input parameters:
\begin{eqnarray}
&& \sqrt{s}=10.58\;{\rm GeV},\quad m_c=1.68\;{\rm GeV}, \quad \alpha(\sqrt{s})=1/130.9,\quad
\alpha_s(\sqrt{s}/2)=0.209.
\end{eqnarray}
The charm pole mass and the running QCD coupling constant are evaluated to two-loop accuracy
with the aid of the package \texttt{RunDec3}~\cite{Herren:2017osy}.

In Table~\ref{Table-SDC:C:one:two:loop} we enumerate the ${\cal O}(\alpha_s)$ and ${\cal O}(\alpha^2_s)$ contributions to the SDCs,
$c_{\lambda_1,\lambda_2}^{J (1,2)}$ for each helicity configuration introduced in \eqref{SDCs:power:series:in:alphas}.
The terms labeled with subscripts $\text{lbl},c$ and $\text{lbl},b$ denote the contributions from the ``light-by-light" diagrams due to charm and bottom loop, 
as illustrated by some typical diagram in Figure~\ref{feynman:diagrams}. We do not include the ``light-by-light" contribution due to light quark loops, since
the net contribution vanishes after including three light quark flavors.
From Table~\ref{Table-SDC:C:one:two:loop}, we observe that the ``light-by-light" contributions are 
insignificant relative to the regular part.

To predict the exclusive production of $J/\psi+\chi_{cJ}$ at B factories, we need some knowledge about the nonperturbative LDMEs. As illustrated in (\ref{NRQCD:vac:to:onium:ME:wvfn:at:origin}), these LDMEs can be estimated by the phenomenological wave functions at the origin. We adopt the radial wave functions at the origin (and their first derivatives) for $J/\psi$, $\psi(2S)$ and $\chi_{cJ}$ evaluated from the Buchm\"uller-Tye (BT) potential model~\cite{Eichten:1995ch} to evaluate the corresponding vacuum-to-charmonium NRQCD matrix elements at initial factorization scale $\mu_\Lambda=1\,\mathrm{GeV}$:
\begin{align}
    \left|R_{J/\psi}(0)\right|^2= 0.81\,{\rm GeV}^3,\quad \left|R_{\psi(2S)}(0)\right|^2= 0.529\,{\rm GeV}^3,
\quad \left|R^\prime_{\chi_{cJ}}(0)\right|^2 = 0.075\, {\rm GeV}^5.
\end{align}

The LDMEs is evolved to another factorization scale by solving the RG equation~(\ref{eq-RG}):
 \begin{align}
    \dfrac{\langle\mathcal{O}_{{}^3L_J}(\mu_\Lambda)\rangle}{\langle\mathcal{O}_{{}^3L_J}(\mu_{\Lambda0})\rangle}
           =\exp\left\{\dfrac{4\gamma_{{}^3L_J}}{\beta_0}\left[\dfrac{\alpha_s(\mu_\Lambda)}{\pi}-\dfrac{\alpha_s(\mu_{\Lambda0})}{\pi}\right]\right\}.
\end{align}

\begin{figure}[h]
    \centering
    \includegraphics[width=0.4\textwidth]{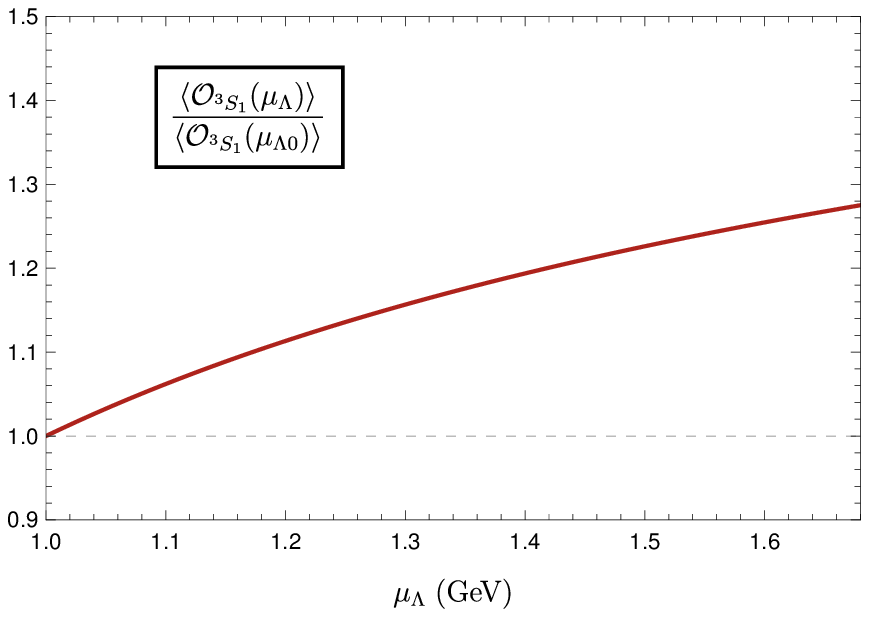}\,
    \includegraphics[width=0.4\textwidth]{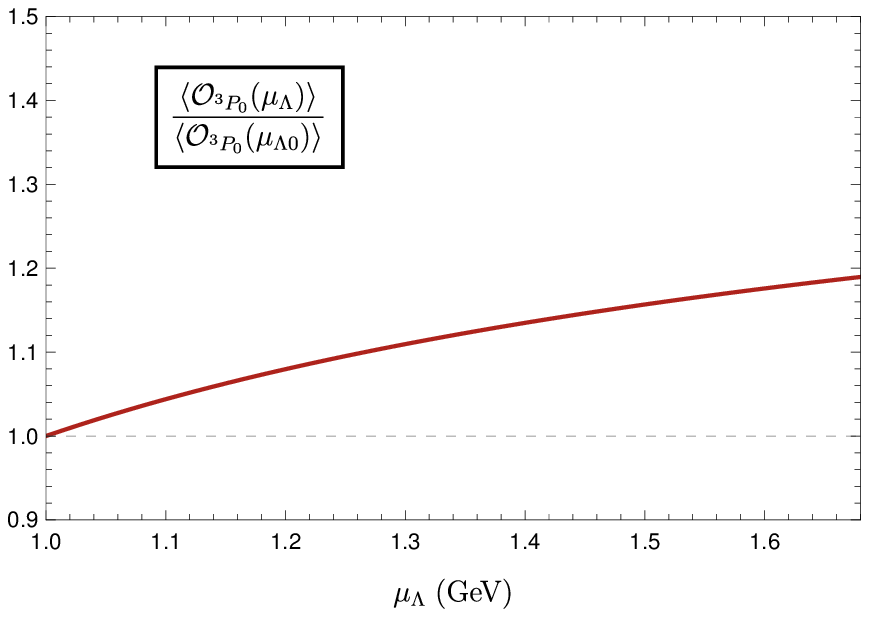}\\
    \includegraphics[width=0.4\textwidth]{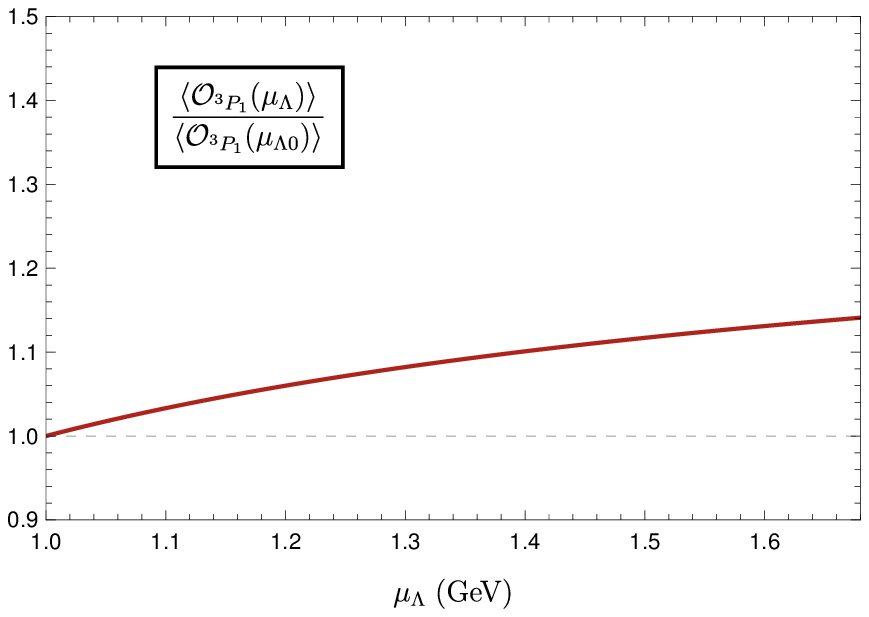}\,
    \includegraphics[width=0.4\textwidth]{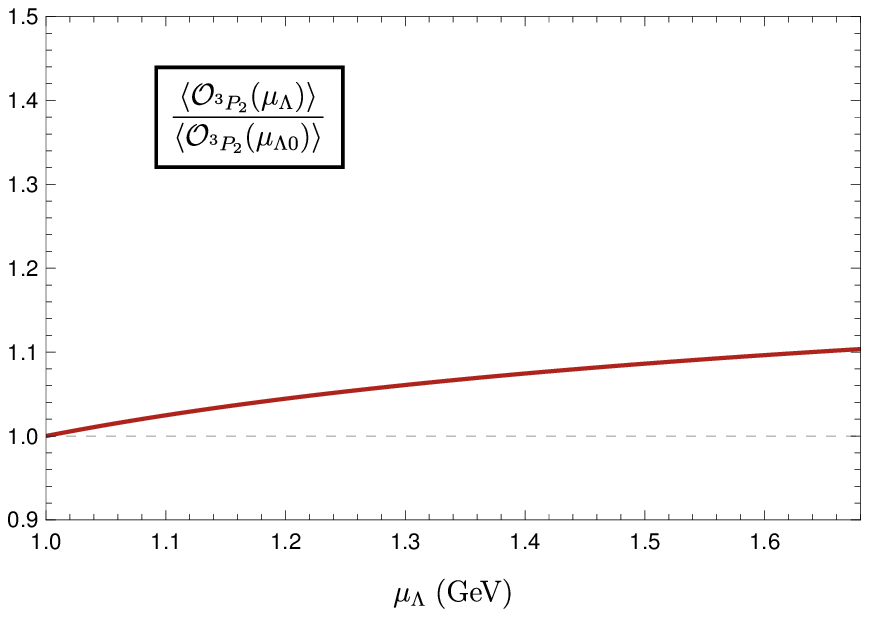}
    \caption{The scale evolution effects of LDMEs. The initial scale $\mu_{\Lambda0}=1$ GeV.}
    \label{fig:LDME muF}
\end{figure}

\begin{figure}
    \centering
    \includegraphics[width=0.4\textwidth]{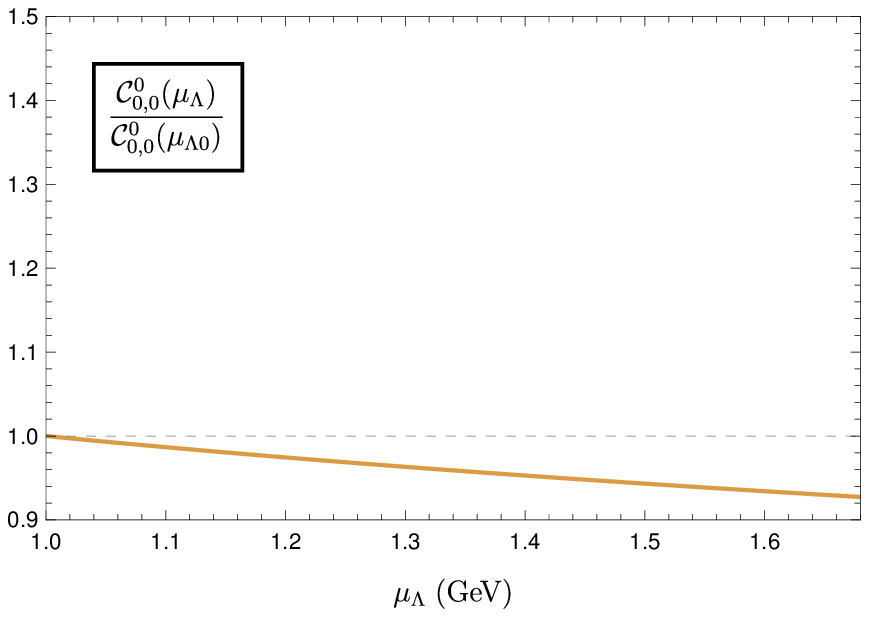}
    \includegraphics[width=0.4\textwidth]{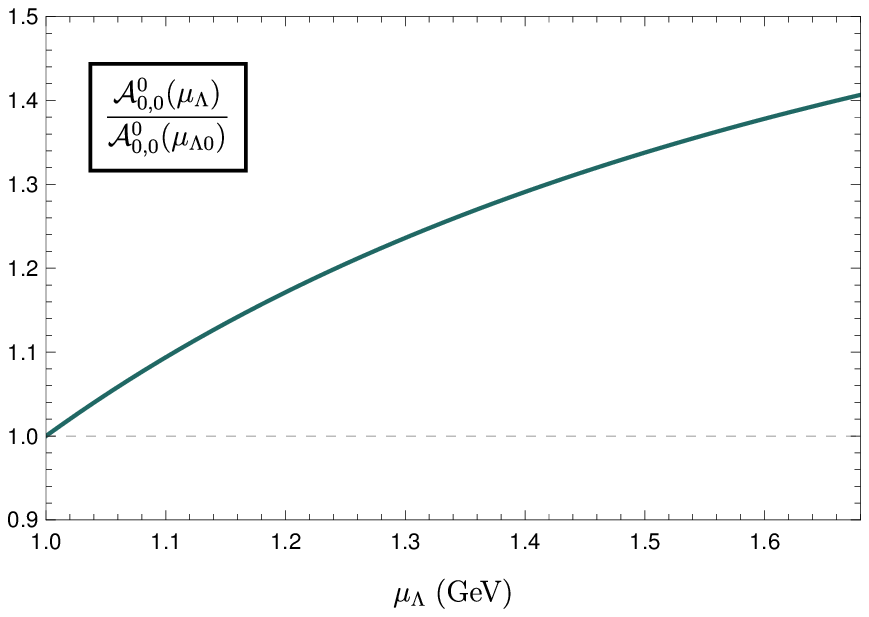}
    \caption{The factorization scale evolution effects of $\mathcal{C}_{0,0}^0$ and $\mathcal{A}_{0,0}^0$. The renormalizaiton sccale is fixed as $\mu_R=\sqrt{s}/2$, and the charm mass is chosen to be $1.68$ GeV. The initial scale $\mu_{\Lambda0}=1$ GeV.}
    \label{fig: muF}
\end{figure}

We plot the scale evolution effects of LDMEs in Figure \ref{fig:LDME muF}, as well as the factorization scale dependence of SDC and helicity amplitude for $J/\psi(0)+\chi_{c0}(0)$ channel in Figure \ref{fig: muF}. We observe that the factorization scale dependences of SDCs and LDMEs tend to cancel. However, since we truncate at $\mathcal{O}\left(\alpha_s^2\right)$, the combined helicity amplitudes still exhibit a large scale dependence.

\begin{table}[tbh]
\caption{Polarized cross sections $\sigma^{(\lambda_1,\lambda_2)}$ (in units of fb) for each helicity channel
$e^+e^-\to J/\psi(\lambda_1)+\chi_{c0,1,2}(\lambda_2)$.
The central values of predictions are obtained by setting $m_c = 1.68\:\text{GeV}$, $\mu_R=\sqrt{s}/2$ and $\mu_\Lambda=1.34\,\mathrm{GeV}$. The first uncertainties is 
given by varying
$\mu_R$ from $2m_c$ to $\sqrt{s}$, and the second uncertainties is evaluated by varying $\mu_\Lambda$ from $1\,\mathrm{GeV}$ to $m_c$. Note that we do not include the uncertainty stemming from the NRQCD matrix elements, which may enhance the predicted cross sections by a factor of $4$. 
\label{Table:Polarized:X:NNLO}}
 \begin{center}
    \resizebox{\textwidth}{!}{
      \begin{tabular}{ccccccc}
          \hline
              &   & $\sigma^{(0,0)}$ & $\sigma^{(1,0)}$ & $\sigma^{(0,1)} (\times 10^{-1})$ &
            $\sigma^{(1,1)} (\times 10^{-2})$ & $\sigma^{(1,2)} (\times 10^{-3})$ \\
    
          \hline
            \multirow{3}{*}{$J/\psi+\chi_{c0}$}
              & LO      
              & $0.79^{+0.27}_{-0.23}$ 
              & $1.28^{+0.44}_{-0.38}$ 
              & -- & -- & -- \\
              & NLO  
              & $1.37^{+0.25}_{-0.26}$ 
              & $2.34^{+0.44}_{-0.45}$ 
              & -- & -- & -- \\
              & NNLO 
              & ${{2.2497}^{+0.0009}_{-0.0800}
              {}^{+0.5534}_{-0.8281}}$ 
              & ${{4.10}^{+0.06}_{-0.25}
              {}^{+1.02}_{-1.52}}$ 
              & -- & -- & -- \\
          \hline
            \multirow{3}{*}{$J/\psi+\chi_{c1}$}
              & LO   & -- 
              & $0.00136^{+0.00046}_{-0.00040}$ 
              & $2.24^{+0.76}_{-0.66}$ 
              & $2.62^{+0.90}_{-0.77}$  & -- \\
              & NLO  & -- 
              & $0.0226^{+0.0204}_{-0.0121}$           
              &$2.72^{+0.20}_{-0.32}$ &
              $2.21^{+0.01}_{-0.10}$ & -- \\
              & NNLO & -- 
              & ${{0.107}^{+0.062}_{-0.045}
              {}^{+0.033}_{-0.042}}$
              & ${{3.13}^{+0.38}_{-0.58}
              {}^{+0.59}_{-1.00}}$ 
              & ${{1.33}^{+0.70}_{-0.61}
              {}^{+0.20}_{-0.39}}$  
              & -- \\
          \hline
            \multirow{3}{*}{$J/\psi+\chi_{c2}$}
              & LO   & 
              $0.202^{+0.069}_{-0.059}$ 
              & $0.159^{+0.054}_{-0.047}$ 
              & $0.328^{+0.112}_{-0.097}$ 
              & $2.62^{+0.90}_{-0.77}$ 
              & $2.72^{+0.93}_{-0.80}$ \\
              & NLO  
              & $0.204^{+0.006}_{-0.016}$ 
              & $0.187^{+0.012}_{-0.021}$ 
              & $0.384^{+0.023}_{-0.041}$  
              & $3.12^{+0.25}_{-0.36}$ 
              & $3.23^{+0.29}_{-0.39}$ \\
              & NNLO 
              & ${{0.172}^{+0.048}_{-0.057}
              {}^{+0.026}_{-0.048}}$ 
              & ${{0.189}^{+0.032}_{-0.043}
              {}^{+0.031}_{-0.055}}$ 
              & ${{0.541}^{+0.009}_{-0.023}
              {}^{+0.097}_{-0.163}}$ 
              & ${{3.14}^{+0.46}_{-0.59}
              {}^{+0.55}_{-0.93}}$ 
              & ${{4.04}^{+0.19}_{-0.35}
              {}^{+0.74}_{-1.23}}$ \\
          \hline
        \end{tabular}
  }
\end{center}
\end{table}

\begin{table}[htb]
\caption{Comparison between our predictions to the unpolarized cross sections and
the measurements in two $B$ factories (in units of fb). The sources of theoretical
uncertainties are the same as in Table~\ref{Table:Polarized:X:NNLO},
respectively. The experimental data are the double charmonium cross sections multiplied by the
branching fractions of $\chi_{cJ}$ decay into more than 2 charged tracks.
The \texttt{Belle} data for $e^+e^-\to \psi(2S) +\chi_{cJ}$ production correspond to
$\chi_{cJ}$ decay into at least 1 charged track~\cite{Belle:2004abn}.}
\label{Table:Unpolarized:X:Section:Comparison}
\begin{center}
    \resizebox{\textwidth}{!}{
              \begin{tabular}{cccc|ccc}
              \hline
                & \multirow{2}{*}{LO} & \multirow{2}{*}{NLO} & \multirow{2}{*}{NNLO} &
                \texttt{Belle} & \textsc{BaBar} \\
                & & & &
                $\sigma \times \mathcal{B}_{>2(0)}$\cite{Belle:2004abn} &
                $\sigma \times \mathcal{B}_{>2}$\cite{BaBar:2005nic} \\
        
              \hline
                $\sigma(J/\psi +\chi_{c0})$ &
                $3.35^{+1.14}_{-0.99}$      & $6.05^{+1.13}_{-1.17}$ 
                & ${{10.45}^{+0.11}_{-0.58}
                {}^{+2.60}_{-3.87}}$
                &$6.4 \pm 1.7 \pm 1.0$ & $10.3 \pm 2.5 ^{+1.4}_{-1.8}$ \\
        
                $\sigma(J/\psi +\chi_{c1})$ &
                $0.503^{+0.172}_{-0.148}$ & $0.63^{+0.08}_{-0.09}$ 
                & ${{0.867}^{+0.006}_{-0.005}
                {}^{+0.188}_{-0.291}}$ 
                &-- & -- \\
        
                $\sigma(J/\psi +\chi_{c2})$ &
                $0.64^{+0.22}_{-0.19}$      & $0.72^{+0.04}_{-0.07}$ 
                & ${{0.728}^{+0.123}_{-0.161}
                {}^{+0.121}_{-0.212}}$  &
                -- & -- \\
        
                $\sigma(J/\psi +\chi_{c1})+\sigma(J/\psi +\chi_{c2})$ &
                $1.15^{+0.39}_{-0.34}$      & $1.36^{+0.12}_{-0.16}$ 
                & ${{1.60}^{+0.12}_{-0.17}
                {}^{+0.31}_{-0.50}}$       &
                $<$5.3 at 90\% C.L. & -- \\
        
                $\sigma(\psi(2S) +\chi_{c0})$ &
                $2.19^{+0.75}_{-0.64}$      
                & $3.95^{+0.74}_{-0.76}$      
                & ${6.82^{+0.07}_{-0.38}
                {}^{+1.70}_{-2.52}}$      
                &$12.5 \pm 3.8 \pm 3.1$ & -- \\
        
                $\sigma(\psi(2S) +\chi_{c1})$ &
                $0.328^{+0.112}_{-0.097}$ 
                & $0.413^{+0.052}_{-0.058}$ 
                & ${0.566^{+0.004}_{-0.003}
                {}^{+0.123}_{-0.190}}$ 
                &
                -- & -- \\
        
                $\sigma(\psi(2S) +\chi_{c2})$ &
                $0.420^{+0.144}_{-0.124}$ 
                & $0.473^{+0.026}_{-0.048}$ 
                & ${0.476^{+0.080}_{-0.105}
                {}^{+0.079}_{-0.14}}$ &
                -- & -- \\
        
                $\sigma(\psi(2S) +\chi_{c1})+\sigma(\psi(2S) +\chi_{c2})$ &
                $0.75^{+0.26}_{-0.22}$      
                & $0.89^{+0.08}_{-0.11}$      
                & ${1.04^{+0.08}_{-0.11}
                {}^{+0.20}_{-0.33}}$      &
                $<$8.6 at 90\% C.L. & -- \\
              \hline
            \end{tabular}
  }
\end{center}
\end{table}

\begin{figure}
\begin{center}
\includegraphics[scale=0.8]{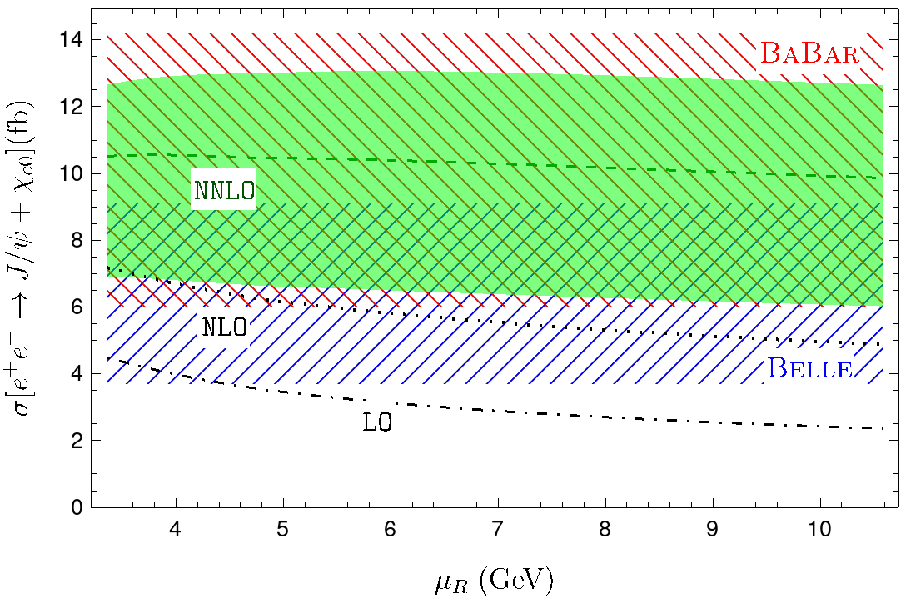}
\includegraphics[scale=0.8]{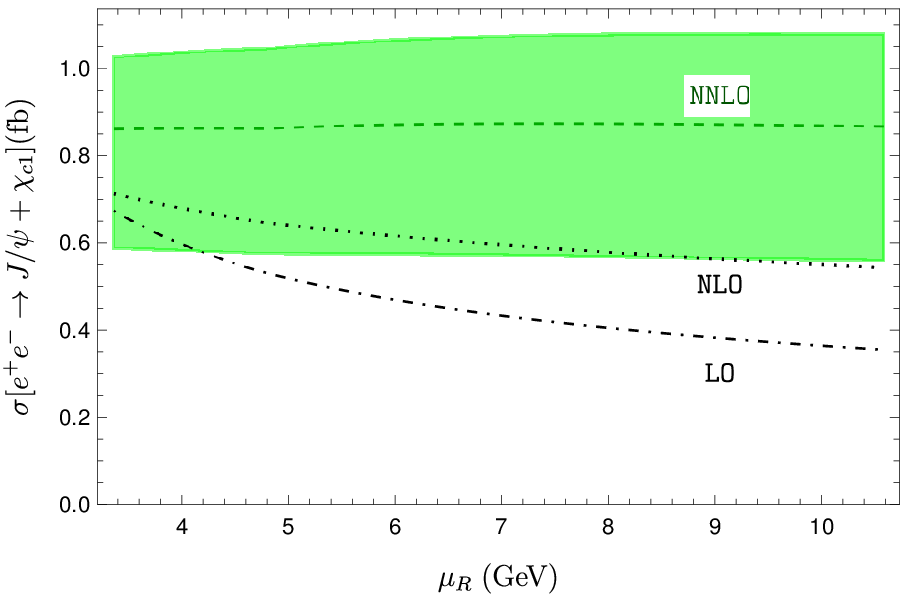}
\includegraphics[scale=0.8]{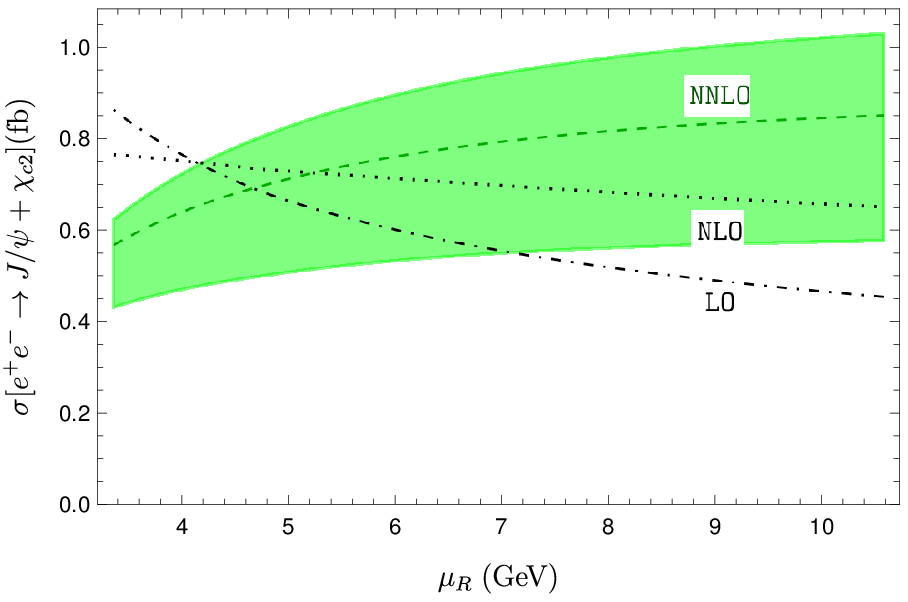}
\includegraphics[scale=0.8]{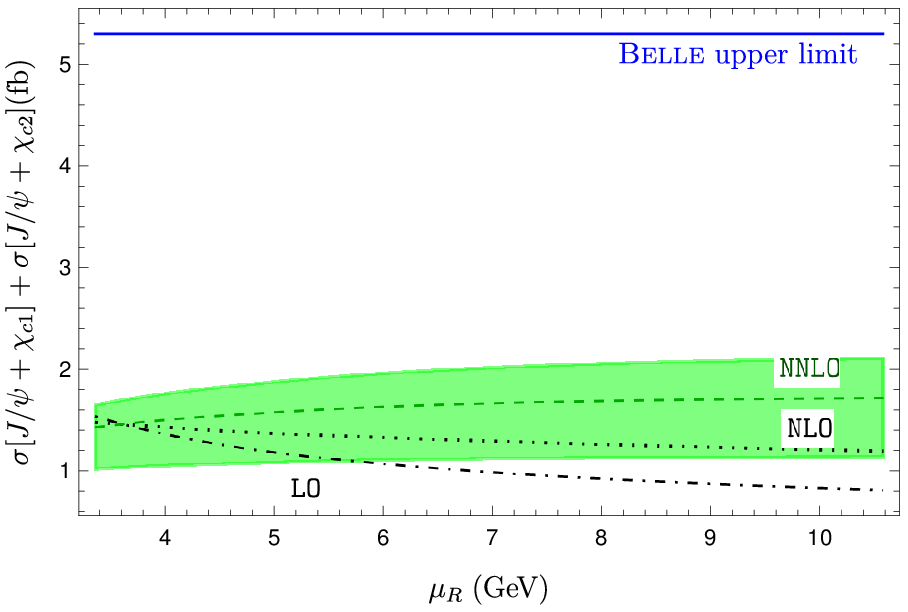}
\caption{NRQCD predictions for the unpolarized cross sections of $e^+e^-\to J/\psi+\chi_{cJ}$ as a function of $\mu_R$ at various level of
 perturbative accuracy. For the sake of clarity, we also juxtapose the $B$ factory measurement for the $J/\psi+\chi_{c0}$ production rate
and the upper bound for combined production rate for $J/\psi+\chi_{c1}$ and $J/\psi+\chi_{c2}$.
\label{Fig:X:Section:Func:of:muR}}
\end{center}
\end{figure}

In Table~\ref{Table:Polarized:X:NNLO} and \ref{Table:Unpolarized:X:Section:Comparison},
we enumerate the NRQCD predictions at various level of perturbative accuracy for the polarized and unpolarized cross sections,  
where we have used the symbols `LO', `NLO', and `NNLO' to denote the computation accurate up to the leading order, the next-to-leading order, and the next-to-next-to leading order in $\alpha_s$ respectively. 
In Table~\ref{Table:Unpolarized:X:Section:Comparison} we also confront our most refined NRQCD predictions
for unpolarized $J/\psi+\chi_{cJ}$ cross sections with the existing $B$ factories measurements.
In Figure~\ref{Fig:X:Section:Func:of:muR}, we also plot the NRQCD predictions for the unpolarized production rates of
$e^+e^-\to J/\psi+\chi_{cJ}$ as function of $\mu_R$ at various levels of perturbative order.
To estimate theoretical uncertainties, we 
vary the renormalization scale $\mu_R$ from $2m_c$ to $\sqrt{s}$ (the first uncertainties)
(taking $\alpha_s(\sqrt{s})=0.176$, $\alpha_s(2m_c)=0.242$ by using \texttt{RunDec3}~\cite{Herren:2017osy}). 
For NNLO predictions, we also slide the factorization scale from $1$ GeV to $m_c$ 
(the second uncertainties),
with the central values obtained by taking
$\mu_R=\sqrt{s}/2$
and $\mu_\Lambda=(1+1.68)\,\mathrm{GeV}/2=1.34$ GeV.
We should emphasize that there are other sources of uncertainties for the predicted cross sections. 
One is the values of the Schr\"odinger wave functions, i.e.,
$\left|R_{J/\psi}(0)\right|^2$ can range from $0.4$ to  $1.5\:{\rm GeV^3}$,
and $\left|R'_{\chi_c}(0)\right|^2$ can range from $0.03$ to $0.13\:{\rm GeV^5}$  in Refs.~\cite{Eichten:1995ch, Chung:2020zqc, Azhothkaran:2020ipl, Radford:2007vd, Choe:2003wx, Gray:2005ur, Chung:2021efj}, which may change the central value of the cross section
by roughly a factor of 4. In addition, the uncalculated relativistic corrections may give arise of $\mathcal{O}(v^2)~\sim 30\%$ corrections to the cross sections.

Examining Table~\ref{Table:Polarized:X:NNLO} closely, one may feel that the polarized cross sections do
not obey the hierarchy as indicated by the helicity scaling rule in \eqref{Polarized:X:Section:Helicity:scaling}.
For example, the cross section in $(1,0)$ channel is nearly twice as large as that in $(0,0)$ channel for $J/\psi+\chi_{c0}$ production,
even after including the ${\cal O}(\alpha^2_s)$ correction. A partial reason might be due to the $B$ factory energy is far from
asymptotically high, so that one may not trust too much on helicity scaling rule.
A strange pattern may be worth comment. The ${\cal O}(\alpha_s)$ and ${\cal O}(\alpha^2_s)$ corrections to $\sigma^{(\pm 1,0)}$ in $J/\psi+\chi_{c1}$ production
appears to be unusually large. 
This might be partially traced to the fact the LO prediction $\sigma^{(0,1)}$ accidently
receives an extra suppression than what is anticipated from the helicity scaling rule (see \eqref{SDC:helicity:amplitude:tree}).
But the helicity scaling rule may be restored  after including higher-order perturbative corrections.

From  Table~\ref{Table:Polarized:X:NNLO} and \ref{Table:Unpolarized:X:Section:Comparison}, we observe that
the ${\mathcal O}(\alpha_s^2)$ corrections have a pronounced impact,
which increase the NLO predictions for most polarized double charmonium production rates and  unpolarized cross sections.

From Table~\ref{Table:Unpolarized:X:Section:Comparison} and Figure~\ref{Fig:X:Section:Func:of:muR}, we also observe that,
compared with the NLO predictions, the renormalization scale dependence in NNLO predictions
are considerably reduced for $\sigma(J/\psi+\chi_{c0,1})$, but slightly worsen for the $J/\psi+\chi_{c2}$ case.

It is also interesting to note that, after incorporating the ${\cal O}(\alpha_s^2)$ corrections, the
NRQCD prediction for the $J/\psi+\chi_{c0}$ production are well consistent with both \texttt{Belle} and \texttt{BaBar} measurements within errors,
and our predictions for the combined production rates of $J/\psi+\chi_{c1}$ and $J/\psi+\chi_{c2}$ is also compatible with
the upper bound placed by \texttt{Belle}. Although the predicted $\sigma(J/\psi+\chi_{c1,2})$ from NRQCD is about one order of magnitude smaller than
$\sigma(J/\psi+\chi_{c0})$, with much greater integrated luminosity, we hope that future \texttt{Belle 2} experiment
will ultimately observe the $e^+e^-\to J/\psi+\chi_{c1,2}$ processes.
It will provide a more critical and comprehensive examination of NRQCD factorization approach.

It is somewhat mysterious that \texttt{Belle} measurement for $\sigma(\psi(2S)+\chi_{c0})$
is twice as large as that for $\sigma(J/\psi+\chi_{c0})$. It is quite difficult to understand this pattern within NRQCD framework,
since the wave function at the origin for $\psi(2S)$ is smaller than that for $J/\psi$.
Concerning quite large experimental uncertainty of \texttt{Belle} data,
we urge future \texttt{Belle 2} experiment to conduct a more accurate measurement
to clarify this confusing situation.

\begin{table}
\caption{NRQCD predictions for the angular distribution parameter $\alpha_J$ (defined in \eqref{Unpolar:Diff:X:Section:Parametrization})
at various perturbative accuracy. The source of the theoretical uncertainties is the same as in
Table~\ref{Table:Polarized:X:NNLO} and \ref{Table:Unpolarized:X:Section:Comparison}.
\label{Table:Angular:Distribution:Parameter} It is worth noting that the value of $\alpha_J$ is insensitive to choice of the NRQCD matrix elements.}
\begin{center}
         \begin{tabular}{cccc|c}
        \hline
          & LO & NLO & NNLO & \texttt{Belle} \\
    
        \hline
          $J/\psi +\chi_{c0}$ 
          & $0.252$                   
          & $0.260^{+0.005}_{-0.004}$
          & ${{0.291}^{+0.014}_{-0.012}
          {}^{+0.002}_{-0.002}}$                       
          &   $-1.01^{+0.38}_{-0.33}$ \\
          $J/\psi +\chi_{c1}$ 
          & $0.697$  
          & $0.739^{+0.028}_{-0.027}$                    
          & ${{0.880}^{+0.054}_{-0.060}
          {}^{+0.004}_{-0.008}}$                       & --- \\
          $J/\psi +\chi_{c2}$ 
          &
          ${-0.197}$
          & ${-0.075}^{+0.012}_{-0.014}$
          & ${{0.025}^{+0.070}_{-0.047}
          {}^{+0.005}_{-0.006}}$ 
          & --- \\
        \hline
      \end{tabular}
\end{center}
\end{table}

In Table~\ref{Table:Angular:Distribution:Parameter} we compare the NRQCD predictions for the angular distribution parameter $\alpha_J$ with the
$B$ factory measurement. From \eqref{Unpolar:Diff:X:Section:Parametrization} to \eqref{Unpolar:Diff:X:Section:Parametrization:chic2},
we note that defined as the ratio of different combinations of the helicity amplitudes, $\alpha_J$ is insensitive to the
nonperturbative NRQCD matrix element. Theoretical uncertainties due to charm mass and renormalization scale appears to be marginal for the
$J/\psi+\chi_{c0,1}$ channels, but becomes substantial for the $J/\psi+\chi_{c2}$ channel.
We observe that, after incorporating higher-order perturbative corrections, the $\alpha_0$ predicted
from NRQCD seems to severely disagree with the \textsc{Belle} measurement.
Actually, even though we consider the contributions from the relativistic corrections $\mathcal{O}(v^2)\sim 30\%$, the theoretical prediction for $\alpha_0$
is still far from the experiment.
From \eqref{Unpolar:Diff:X:Section:Parametrization:chic0}, one tells the \texttt{Belle} measurement indicates that
the $J/\psi+\chi_{c0}$ production is dominated by the helicity-conserving $(0,0)$ channel, which seems compatible with helicity scaling law.
Nevertheless,  Table~\ref{Table:Polarized:X:NNLO} indicates that the pattern is drastically opposite in NRQCD,
in which $\sigma^{(\pm 1,0)}$ is about twice bigger than $\sigma^{(0,0)}$! Needless to say, hopefully we have to wait for the
future \texttt{Belle 2} experiment to settle this disquieting discrepancy.

\section{Summary\label{Summary}}

Within the  framework of NRQCD factorization, we compute the $\mathcal{O}(\alpha_s^2)$ perturbative corrections to
$e^+e^-\to J/\psi+\chi_{cJ}$ ($J=0,1,2$) production at (super) $B$ factory.
With the aid of the newly developed AMF method, we are able to present
the (un)polarized cross section and $J/\psi$ angular distribution through order-$\alpha_s^2$ with high numerical accuracy.
At $\mathcal{O}(\alpha_s^2)$, we observe that the renormalization scale dependence for $\sigma(J/\psi+\chi_{c0,1})$ are significantly reduced,
while get slightly worsen for $\sigma(J/\psi+\chi_{c2})$. Our theoretical predictions are quite sensitive to the choice of charm quark mass.
Approximating the NRQCD matrix elements by the (derivative of) wave functions at the origin in the potential model,
our most refined prediction is $\sigma(J/\psi+\chi_{c0})={{10.45}^{+0.11}_{-0.58}{}^{+2.60}_{-3.87}}$ fb,
where the first uncertainty is estimated by varying renormalization scale and the second uncertainty originates
from sliding the NRQCD factorization scale\footnote{In this work, we do not include the uncertainty inherent in the NRQCD matrix elements, which may bring significant uncertainty for the predicted cross sections. Fortunately, 
as the ratios of linear combination of squared helicity amplitudes, 
the predicted angular distribution parameters $\alpha_J$ are insensitive to the values of the NRQCD matrix elements. }. 
This prediction is consistent with two $B$ factory measurements
within uncertainties. Our predictions at two-loop accuracy for
$\sigma(J/\psi+\chi_{c1,2})$ are about one order of magnitude smaller than $\sigma(J/\psi+\chi_{c0})$,  which are
compatible with the upper limit of
the \texttt{Belle} measurement, and will likely be observed at future \texttt{Belle 2} experiment.
On the other hand, we find there also emerges severe discrepancy between the most refined NRQCD predictions and the measurements.
One example is the total cross section for $e^+e^-\to \psi(2S)+\chi_{c0}$.
The other example is the angular distribution parameter for $e^+e^-\to \psi(2S)+\chi_{c0}$.
Our prediction $\alpha_0(\chi_{c0})={{0.291}^{+0.014}_{-0.012}
          {}^{+0.002}_{-0.002}}$ is in sheer contradiction to the  measured value
$-1.01^{+0.38}_{-0.33}$ by \texttt{Belle}.
Settling down these discrepancies calls for more theoretical and experimental efforts.
We hope that future \texttt{Belle 2} experiment will shed crucial light on the mechanism of
exclusive double charmonium production and the applicability of NRQCD factorization.

\section*{Acknowledgments}

The work of W.-L. S. is
supported by the National
Natural Science Foundation of China under Grant No. 11975187. 
The work of F.~F. is supported by the National Natural
Science Foundation of China under Grants No. 12275353,
  No. 11875318.
The work of Y.~J., Z.-W.~Mo. and J.-Y.~Z. is supported in part by the National Natural Science Foundation of China under Grants No. 11925506 and
No. 12070131001 (CRC110 by DFG and NSFC).
This work was supported in part by the Natural Science Foundation of China under Grant No.11847301 and by the Fundamental Research Funds for the Central Universities under Grant No. 2019CDJDWL0005









\end{document}